# Low-temperature thermal expansion of rock-salt ZnO

Petr S. Sokolov,[1] Andrey N. Baranov,[2] Anthony M.T. Bell,[3] and Vladimir L. Solozhenko [1,*]

[1] *LSPM–CNRS, Université Paris Nord, 93430 Villetaneuse, France*

[2] *Chemistry Department, Moscow State University, Moscow 119991, Russia*

[3] *HASYLAB–DESY, 22603 Hamburg, Germany*

Lattice parameter of metastable high-pressure phase of zinc oxide, rock-salt ZnO was measured in the 10–300 K temperature range using synchrotron X-ray powder diffraction. No phase transition was observed down to 10 K. The lattice parameter of rock-salt ZnO was found to increase from 4.266 Å in the 10–80 K range up to 4.2752(3) Å at 298 K, while the volume thermal expansion coefficient increases from slight negative values below 40 K up to $4.77 \times 10^{-5}$ K$^{-1}$ at 298 K.

Zinc oxide belongs to the family of wide-band-gap semiconductors with strong ionic character of chemical bonds. At ambient conditions ZnO has wurtzite structure (P6$_3$mc, *w*-ZnO) that transforms into rock-salt one (Fm3m, *rs*-ZnO) at pressures above 5 GPa.[1,2] Upon pressure release *rs*-ZnO reverts back to wurtzite phase.[3] Very recently we have shown that nanocrystalline *rs*-ZnO synthesized at high pressures and high temperatures can be completely recovered at normal conditions.[2] Here we report for the first time the low-temperature thermal expansion of metastable *rs*-ZnO at ambient pressure.

Single-phase nanocrystalline bulk *rs*-ZnO has been synthesized from *w*-ZnO nanopowder (grain size of ~9 nm) at 7.7 GPa & 800 K and subsequent rapid quenching.[2] The low-temperature (10–300 K) thermal expansion of the recovered *rs*-ZnO was studied at the B2 powder diffraction beamline of the DORIS III storage ring (HASYLAB–DESY).[4] Debye-Scherrer geometry with rotating borosilicate glass capillary was used. The X-ray diffraction patterns were collected in the 2–70° 2Θ-range (λ = 0.69797 Å) for 10 min in real time using the OBI image plate detector.[5] The temperature in the helium closed-cycle cryostat was kept constant within 1 K using a PID controller and silicon diode temperature sensor. The measurement cycle started at 10 K. Each following temperature point was set by increasing the heater power. Before each data collection the sample temperature was stabilized for 5 minutes. The diffraction data were analyzed using DatLab software, and positions of diffraction lines were determined by fitting to the Pearson profile function. Grain size and lattice strain of *rs*-ZnO have been calculated from the width of X-ray diffraction lines using Williamson-Hall method.[6] Direct measurement of *rs*-ZnO grain size has been performed by transmission electron microscopy (TEM) using JEOL 2011 microscope. The micro-Raman spectra were recorded using a Renishaw InVia Raman microscope equipped with 514 nm argon laser (5 μm beam) and Oxford MG11 continuous flow helium cryostat for low-temperature (10–300 K) measurements.

---

* Corresponding author: e-mail: vladimir.solozhenko@univ-paris13.fr

At room temperature, the lattice parameter of recovered nanocrystalline $rs$-ZnO (grain size is 58(8) nm with lattice strain of 0.19(2)% according to X-ray studies and 60–90 nm according to TEM) makes 4.2752(3) Å. Down to 10 K, the sample remains rock-salt structure and does not show any tendency for phase transition. The X-ray diffraction pattern of $rs$-ZnO sample taken at 10 K is presented in Fig. 1. After low-temperature measurements, the 298-K lattice parameter of the sample was found to be 4.2749(3) Å. The total absence of characteristic $w$-ZnO modes in Raman spectra at room and low (down to 10 K) temperatures is in perfect agreement with X-ray diffraction data that additionally proves the phase purity and homogeneity of the sample.

The experimental data on $rs$-ZnO lattice parameter versus temperature are shown in Fig. 2. The lattice parameter remains almost constant below 80 K, above this temperature lattice expansion becomes pronounced. In order to find analytical expression best describing our $a$ versus T data, we fitted them to several polynomial functions, from second to fifth degree. The criterion for the best fit was comparison of the root-mean square of the residuals to the experimental uncertainty. Finally, for approximation of the experimental data in the 10-300 K range we have chosen the second-degree polynomial similar to that in Ref. 7. It should be noted that the study of thermal expansion at very low (below 40 K) temperatures requires special equipment and long (~10 cm) single-crystal samples,[8] which is impossible in the case of metastable phases recovered from high pressure and high temperature; $rs$-ZnO, in particular.

Thus, the experimental data ($a$ in Å, T in K) were fitted to a second-degree polynomial (red line in Fig. 2):

$$a = 4.26617(9) \times [1 - 2.49(36) \times 10^{-6} \times T + 3.09(12) \times 10^{-8} \times T^2] \qquad (1)$$

The lattice parameters calculated using equation (1) are very close the corresponding experimental values (see Table 1), showing the good quality of the fit ($R^2 = 0.99241$).

The values of linear thermal expansion coefficient were calculated by differentiating the equation (1); α is linear with temperature and can be expressed by equation

$$\alpha = -2.48(2) \times 10^{-6} + 6.1705(11) \times 10^{-8} \times T \qquad (2)$$

The corresponding volume thermal expansion coefficient $\beta = 3 \times \alpha$ increases from $-0.13 \times 10^{-5}$ K$^{-1}$ at 20 K to $4.77 \times 10^{-5}$ K$^{-1}$ at 298 K which is close to the $5.2(2) \times 10^{-5}$ K$^{-1}$ experimental value reported in Ref. 9 and $4.7 \times 10^{-5}$ K$^{-1}$ value from first-principle calculations.[10] At room temperature thermal expansion coefficient of $rs$-ZnO is about three times higher than that of $w$-ZnO ($1.57 \times 10^{-5}$ K$^{-1}$ according to Ref. 7).

At ambient pressure nanocrystalline *rs*-ZnO is kinetically stable to ~360 K only,[11] hence, the direct experimental study of its high-temperature thermal expansion is not possible. However, extrapolation of the thermal expansion data for metastable cubic ZnO-rich solid solutions $x$ZnO-($1$-$x$)LiMO$_2$ (M = Fe$^{3+}$, Ti$^{3+}$, Sc$^{3+}$)[11,12] and $x$ZnO-($1$-$x$)MeO (Me = Ni$^{2+}$, Co$^{2+}$)[11,13] to $x$ = 1 (pure *rs*-ZnO) gives the almost constant β-value of about 5×10$^{-5}$ K$^{-1}$ in the 300-800 K temperature range.

The shape of the reduced thermal expansion curve of *rs*-ZnO is similar to those of other metal monoxides with NaCl structure[7] i.e. CaO, MgO and BaO (see Fig. 3). This similarity, together with the same temperature range in which the negative expansion coefficients prevail (10–40 K), is evidence for a similar kind of phonon modes for points of high symmetry in q-space for all these structures. The further study is required to understand the phenomenon.

In summary, thermal expansion of rock-salt ZnO, metastable high-pressure phase, has been experimentally studied for the first time in the 10–300 K temperature range at ambient pressure using synchrotron X-ray powder diffraction. The volume thermal expansion coefficient was found to increase from almost zero value in 10–80 K range to 4.77×10$^{-5}$ K$^{-1}$ at 298 K and remains virtually constant (5×10$^{-5}$ K$^{-1}$) at higher temperatures.

The authors are grateful to Dr. O. Brinza for TEM studies and Dr. A.A. Eliseev for low-temperature Raman measurements. X-ray diffraction studies at B2 beamline, DORIS III have been carried out during beamtime allocated to the Project DESY-D-I-20100021 EC at HASYLAB-DESY and have received funding from the European Community's Seventh Framework Programme (FP7/2007-2013) under grant agreement No 226716. This work was partially supported by the Russian Foundation for Basic Research (Project No 11-03-01124). PSS is thankful to the "Science and Engineering for Advanced Materials and devices" (SEAM) Laboratory of Excellence for financial support.

**Figure 1.**

Experimental (circles), calculated (solid line) and difference (lower) X-ray powder diffraction pattern ($\lambda = 0.69797$ Å) of the recovered $rs$-ZnO at 10 K and ambient pressure. Vertical bars indicate the Braggs peak positions for rock-salt ZnO.

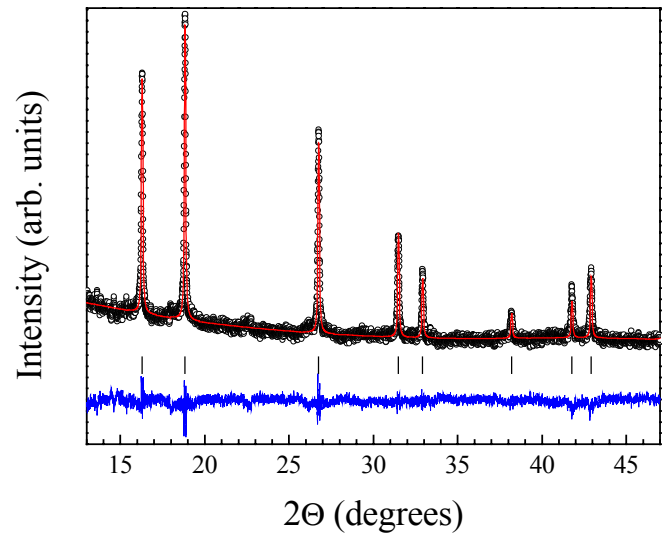

**Figure 2.**

Lattice parameter of rock-salt ZnO versus temperature at ambient pressure.

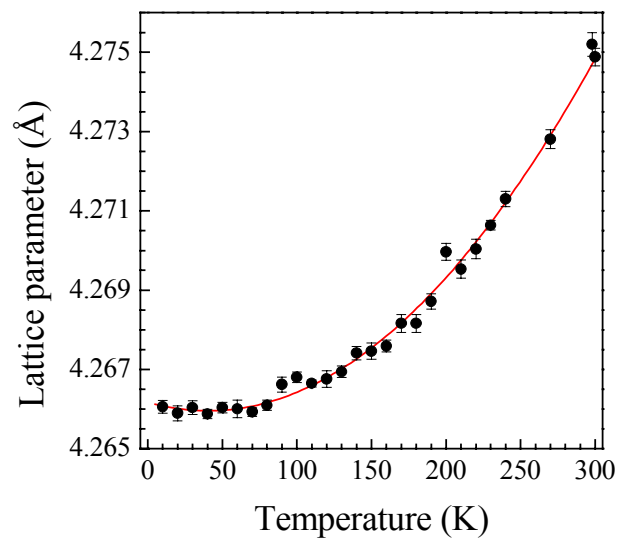

**Figure 3.**

Reduced thermal expansion curves for rock-salt ZnO (present work) and MgO, CaO, BaO [7].

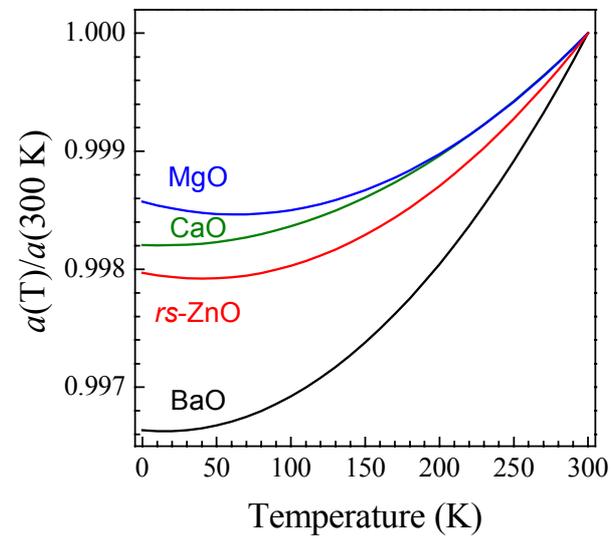

**Table 1.**

Lattice parameter and thermal expansion of rock-salt ZnO.

| T (K) | a (Å) | | α×10$^5$ (K$^{-1}$) |
|---|---|---|---|
| | observed | calculated | |
| 10 | 4.26606(16) | 4.26608 | — |
| 20 | 4.26589(19) | 4.26601 | -0.1254 |
| 30 | 4.26603(17) | 4.26597 | -0.0636 |
| 40 | 4.26588(12) | 4.26596 | -0.0018 |
| 50 | 4.26604(13) | 4.26597 | 0.0600 |
| 60 | 4.26600(18) | 4.26601 | 0.1218 |
| 70 | 4.26592(11) | 4.26607 | 0.1836 |
| 80 | 4.26610(13) | 4.26616 | 0.2454 |
| 90 | 4.26662(19) | 4.26628 | 0.3072 |
| 100 | 4.26680(13) | 4.26643 | 0.3690 |
| 110 | 4.26665(10) | 4.26660 | 0.4308 |
| 120 | 4.26676(20) | 4.26679 | 0.4925 |
| 130 | 4.26694(15) | 4.26702 | 0.5543 |
| 140 | 4.26741(17) | 4.26727 | 0.6160 |
| 150 | 4.26746(21) | 4.26754 | 0.6778 |
| 160 | 4.26759(15) | 4.26785 | 0.7395 |
| 170 | 4.26816(23) | 4.26817 | 0.8012 |
| 180 | 4.26816(23) | 4.26853 | 0.8629 |
| 190 | 4.26872(20) | 4.26891 | 0.9246 |
| 200 | 4.26997(21) | 4.26932 | 0.9863 |
| 210 | 4.26953(23) | 4.26975 | 1.0473 |
| 220 | 4.27004(25) | 4.27021 | 1.1096 |
| 230 | 4.27064(12) | 4.27070 | 1.1712 |
| 240 | 4.27130(19) | 4.27121 | 1.2327 |
| 270 | 4.27281(23) | 4.27291 | 1.4174 |
| 298 | 4.27520(30) | 4.27471 | 1.5895 |
| 300 | 4.27488(25) | 4.27485 | 1.6017 |